\documentstyle{article}    % Tell LaTeX the document type<<<<<<<<<<<<

\def\beq{\begin{equation}}
\def\eeq{\end{equation}}
\def\Omeg{{\mit\Omega}}
\def\Ph{{\mit\Phi}}
\def\Ps{{\mit\Psi}}

 \normalmarginpar
\twocolumn
 \makeatletter
 \gdef\@proofbox{\relax}
 \long\def\proofbox#1{\gdef\@proofbox{#1}}
 \proofbox{\small{\sl PhysComp96}\\\ifx\UNDEF\@fullpaper
     Extended abstract\else Full paper\fi\\Draft, \@date}

 \gdef\fullpaper{\gdef\@fullpaper{}}

 \gdef\@author{John Doe, {\sl No-Name University, Shipping Dept.}}
 \def\author#1{\gdef\@author{#1}}

 \gdef\@abstract{}
 \long\def\abstract#1{\gdef\@abstract{#1}}

\def\@maketitle{\newpage\leavevmode
  \begin{minipage}[t]{0.30\textwidth}
    \hrule height0pt
    \raggedright
    \mbox{}\par
    \@proofbox
  \end{minipage}\relax
  \begin{minipage}[t]{0.70\textwidth}
    \hrule height0pt
    \raggedleft
    \LARGE\@title\par
    \vskip4pt
    \large\@author
  \end{minipage}
  \vskip8pt
  \ifx\@abstract\@empty\else{\vskip.5em%
\leftskip1.5in\parskip4pt\small\@abstract\par\vskip.5em}\fi
  \rule{\textwidth}{0.4pt}
  \vskip16pt}

  \def\@arabic#1{\number #1} % my redefinition

\def\comppad{\thinspace}
\def\comp{\comppad\begingroup \tt \let\do\@makeother \dospecials 
          \@ifstar{\@scomp}{\@comp}}
\def\@scomp#1{\def\@tempa ##1#1{##1\endgroup\comppad}\@tempa}
\def\@comp{\obeyspaces \frenchspacing \@scomp}

\makeatother

 \title{Error symmetrization\\ in quantum computers}
 \author{Asher Peres\\ Technion, Haifa, Israel}
 \date{12 May 1996}

\begin{document}

\maketitle

A computer is a physical system, subject to the ordinary laws of nature.
No error ever occurs in the application of these laws. What we call an
error is a mismatch between what the computer does and what we wanted it
to do. This may be caused by incorrect programming (software errors,
that I shall not consider here), or by imperfect hardware. The computer
engineer's problem is to design the hardware in such a way that common
flaws, which are unavoidable, will almost never cause errors in the
final output (namely, in the relevant parts of the final state of the
computer).

In conventional, classical computers, bistable physical elements are
used for representing logical bits, 0 and 1. Thermal fluctuations (or
other imperfections) may flip such a bistable element, causing an error.
A simple method for reducing errors is the use of redundancy: the result
of a ``majority vote'' in a redundant array is deemed to be correct,
because, if the error probability is small, the probability of the
majority being wrong is exponentially small. A more sophisticated and
more efficient method for error correction is the use of codewords,
which can be unambiguously recognized and corrected after a finite
number of errors \cite{Welsh}.

In quantum computers, logical bits (called qubits) are not restricted to
the discrete values 0 and 1. Their ``value'' (or ``state'') may be
represented by a point on a unit sphere. Moreover, that state may not be
definite, because several qubits may have their states inseparably
entangled: the entire computer has, ideally, a definite quantum state,
but each individual qubit, considered separately, is in an incoherent
mixture of states. The continuous nature of qubit states implies that
there can be no intrinsic stabilizing mechanism, and error control
becomes critical.

Here, a distinction must be made between quantum computers of the
Benioff type \cite{Benioff,Feynman}, where quantum hardware is used for
implementing classical logic, and computers that are fundamentally
quantal \cite{Deutsch}, and can do more than just mimicking classical
computation. In the former case, there are instants of time at which all
the qubits ought to represent definite values, 0 or 1. They are not then
in a quantum superposition, and error correction can be done as for a
classical computer \cite{Peres}. On the other hand, in a computer of the
Deutsch type \cite{Deutsch}, the quantum state of the computer typically
is an entangled state of all the qubits, and classical methods of error
correction are not applicable. What can be done then depends on the
nature of the expected errors.

In general, we may write the Hamiltonian of the computer as $H=H_0+H_1$,
where $H_0$ is the Hamiltonian of an ideal error free computer, and
$H_1$ represents the influence of the environment. The latter is unknown
to the computer designer, except statistically. That Hamiltonian acts
on a Hilbert space which is the tensor product of those representing the
computer and the environment. The designer's problem is to distill, from
the computer's variables, a subset giving with probability close to 1
the correct result of the computation, irrespective of the unknown form
of $H_1$ and of the initial state state of the environment. Two
different types of errors ought to be considered: accidental large
disturbances to isolated qubits (e.g., a residual gas molecule may hit
one of them), and small, random, uncorrelated drifts of all the qubits.

The first type of error can be corrected by using codewords, as first
shown by Shor \cite{Shor}. A codeword is a representation of a logical
qubit by means of several physical qubits. There were 9 qubits in Shor's
codewords. It is now known that the minimal number is 5. In particular,
Bennett {\sl et al.\/} have constructed 5-qubit codewords that have the
remarkable property of being invariant under a cyclic permutation of the
qubits \cite{BDSW}. In all these quantum codewords, the physical qubits
are in a highly entangled state, chosen in such a way that, if any one
of the qubits gets entangled with an unknown environment, there still is
enough information stored in the other qubits to restore the codeword
and to unitarily disentangle it from the environment, irrespective of
the unknown state of the latter. (Some authors use a ``quantum
measurement'' for finding the error syndrome, and performing the
necessary correction.  This is not at all necessary \cite{Peres}: you
don't have to know the error in order to correct it. This can be done
automatically by a unitary transformation. The correcting qubits are
then left in an unknown state and have to be discarded. They cannot be
used again unless they are restored to their initial state by a
dissipative process.)

The second type of error, continuous random drifts of all the qubits,
cannot be eliminated by using codewords, but can be reduced by
symmetrizing the quantum state \cite{BBDEJM,BDJ}. In this paper, I show
how the symmetrization method can be improved and combined with the use
of codewords. That method, in its original version, involved the use of
$R$ identical replicas of the entire computer. At preset times, the
joint quantum state of the $R$ computers is projected onto the symmetric
subspace of their common Hilbert space (for example, by measuring
whether or not the state is symmetric, and aborting the computation if
the answer is negative). It can be shown that, if the error probability
is small, this projection further reduces it by a factor $R$, on the
average. On the other hand, symmetrization gives poor results if a
single qubit goes completely astray, because we then have a
non-symmetric state that is almost orthogonal to the symmetric subspace,
and the computation is almost always aborted. Indeed, if one
of the computers has a state orthogonal to that of all the others, the
probability is only $1/R$ that the joint state will be projected onto
the symmetric subspace, and in that case, the error is not eliminated,
but rather uniformly spread over all the $R$ computers! Obviously, large
isolated errors are best handled by means of codewords.

There is however a more efficient protocol for error correction by
symmetrization. The $R$ computers can be arranged in pairs, and each one
of the $R/2$ pairs symmetrized separately. The process can then be
repeated with different pairing arrangements, if we wish to further
improve the symmetry. With such a pairwise symmetrization, if a computer
accidentally gets into a state orthogonal to that of all the other ones,
there is a 50\% chance that the pair containing the bad computer will be
eliminated, and a 50\% chance that the error will be equally shared by
the two computers. Repeating this process many times, so that each
computer has many partners, ultimately leads to the elimination of the
bad computer, together with one good one. There still are $R-2$ good
computers left.

A more complicated (and perhaps more realistic) model would be to assume
that any computer may occasionally fail when one of the logical steps is
executed. This event must be rare enough so that the total probability
of failure of any given computer during the entire computation is less
than 1/2. Pairwise symmetrizations are performed between any two
logical steps (the pairs are chosen in such a way that each computer is
compared with many other ones during the complete computation). Most
errors are then eliminated, and the surviving computers contain, on
the average, less than one defective result.  In this theoretical model,
an ``error'' means a state that is orthogonal to the correct one. This
has to be generalized to the case of less radical errors. It is
plausible that repeated pairwise symmetrizations are in general
preferable to a single overall symmetrization, but a formal proof is
still needed.

Instead of symmetrizing the combined state of several complete
computers, we may also symmetrize individual codewords, if the latter
have an internal symmetry. For example the codewords of ref.~\cite{BDSW}
are invariant under cyclic permutations of their 5 qubits. These
codewords were designed in such a way that if any four qubits are
correct, it is always possible to restore the remaining defective qubit.
However, the codeword error correction procedure definitely requires
four qubits to be correct, and it cannot cope with small drifts of all
five qubits. Therefore, it is helpful to test once in a while the cyclic
symmetry of the codeword: successful tests will reduce the amplitude of
small errors. Unfortunately, just as in the case of inter-computer
symmetrization, an unsuccessful test leads to an asymmetric state, and
forces us to completely discard the incorrect codeword. One of the
logical qubits is then missing, and the computation can proceed only if
there is enough redundancy among the logical qubits themselves (not only
in their representation by physical qubits), for example, if they
are parts of higher order codewords.

Clearly, the poor efficiency of the symmetrization method is due to
possible failures of the symmetry tests (also known as ``quantum
measurements''). If a test fails, we must discard a codeword, or an
entire computer, or the entire process. However, there is really no need
of a measurement in order to force a quantum state to stay in a
symmetric subspace. A measurement is not a supernatural event. It is an
ordinary dynamical process, and any error correction that may result
from it should also be obtainable as a consequence of a unitary
evolution, governed by ordinary dynamical laws. Indeed, a much simpler
method for enforcing symmetry of the quantum state is to impose on the
$R$ computers an extra static potential that vanishes in the symmetric
subspace, and has a very large value in all the orthogonal (asymmetric)
states. Effectively, in the $R$ computers, any $R$ homologous physical
qubits behave as if they were $R$ bosons. Likewise, if the 5 qubits of a
codeword have cyclic symmetry \cite{BDSW}, we may protect their cyclic
subspace by erecting around it a high potential barrier. The result of
such a symmetrizing potential is analogous to a continuous Zeno effect
\cite{qt}.

As a simple quantitative example, consider two computer memories,
each one consisting of a single qubit, initially in the state
$\alpha\choose\beta$. I am using here the terminology and notations
appropriate to spin-$1\over2$ particles. A symmetric state of the pair
belongs to the triplet $(J=1)$ representation, while the singlet
$(J=0)$ is antisymmetric. We want these computer memories to be stable:
there should be no evolution of the two qubits. The problem is to
protect them against random fluctuations of the environment. Let us use
for this purpose a Hamiltonian,

\beq H_0=(1-{\bf J}^2/2)\,\Omeg, \eeq
where $\Omeg$ is a large positive constant. Since ${\bf J}^2=J(J+1)$,
this potential vanishes in the triplet state, and is equal to $\Omeg$
for a singlet. As a simple model of perturbation, let

\beq H_1=a\,\sigma_{Az}+b\,\sigma_{Bz}, \eeq
where $a$ and $b$ are constant coefficients much smaller than $\Omeg$,
and the subscripts $A$ and $B$ refer to the two qubits. This can also be
written as

\beq H_1=\epsilon\,(\sigma_{Az}+\sigma_{Bz})+
 \eta\,(\sigma_{Az}-\sigma_{Bz}), \eeq
where $\epsilon=(a+b)/2$ and $\eta=(a-b)/2$. The $\epsilon$ term in
$H_1$ is symmetric, it commutes with $H_0$, and therefore this kind of
perturbation cannot be eliminated by symmetrization. Indeed, the
evolution of the qubit state $\alpha\choose\beta$ is given (if we ignore
the $\eta$ term, for simplicity) by $\alpha(t)= \alpha(0)\,e^{-i\epsilon
t}$ and $\beta(t)=\beta(0)\,e^{i\epsilon t}$. If there were $R$ qubits,
instead of just two, the symmetric part of the perturbation (which
cannot be eliminated by symmetrization) would have as its coefficient
the arithmetic average of the individual perturbations. If the latter
are random and independent, that average is expected to be smaller than
the rms perturbation by a factor $\sqrt{R}$. No further reduction can be
expected.

On the other hand, the error due to the antisymmetric part of $H_1$ can
be considerably reduced. Written with the Bell basis \cite{BMR}, the
initial state of the pair is

\beq {\alpha\choose\beta}\otimes{\alpha\choose\beta}=
 \alpha^2\,\Ph^+ +\beta^2\,\Ph^- + 2\alpha\beta\,\Ps^+.\label{state}\eeq
The antisymmetric part of the perturbation has matrix elements given by

\beq (\sigma_{Az}-\sigma_{Bz})\,\Ph^\pm=0, \eeq
\beq (\sigma_{Az}-\sigma_{Bz})\,\Ps^\pm=\Ps^\mp. \eeq
The nontrivial part of the Hamiltonian thus involves only the $\Ps^\pm$
subspace. We can write (ignoring for simplicity the $\epsilon$
contribution, which is symmetric)

\beq H=H_0+H_1= \left( \begin{array}{cc} 0 & \eta \\
 \eta & \Omeg \end{array}  \right). \eeq
It is easy to find the eigenvalues and eigenvectors of this Hamiltonian.
The initial state (\ref{state}) can be written as a linear combination
of these two eigenstates, and its time evolution be obtained:
the $\Ph^\pm$ terms have constant amplitudes, and, for $\eta\ll\Omeg$,
the $\Ps^+$ term in (\ref{state}) evolves as

\beq \Ps^+\to e^{i\eta^2t/\Omeg}\,\Ps^+
 +(\eta/\Omeg)\,(e^{-i\Omeg t}-1)\,\Ps^-, \eeq
where terms of order $(\eta/\Omeg)^2$ have been neglected. If we can
make $\Omeg$ arbitrarily large (as we do in an ideal ``quantum
measurement'' context, where the interaction with the measuring
apparatus is assumed arbitrarily strong), the $\Ps^+$ term is perfectly
stabilized. For large but finite $\Omeg$, the amplitude of the $\Ps^-$
term is always small, but the $\Ps^+$ term undergoes a slow secular
drift, which definitely is an error, but is compatible with the symmetry
constraint. The same kind of drift also occurs for repeated discrete
symmetrization \cite{BBDEJM,BDJ}, because the symmetric state obtained
at each step contains a small residual error, and these errors gradually
accumulate.

These considerations can now be generalized from 2 to $R$ computers,
each one having many qubits. It may seem that a global potential is
required, involving all of them at once, a proposal that would be a
technological nightmare. Fortunately, this is not necessary: it is
enough to take $R(R-1)/2$ identical potentials, one for each pair of
computers. If any two computers are in a symmetric state, then all $R$
computers are in a symmetric state, by definition. Moreover, the
comparison of any two computers can be done bitwise: if each one has $L$
qubits, one needs $L$ two-qubit potentials, that vanish for the triplet
state, and are very large for the singlet state. How to actually realize
this two-qubit interaction depends on how the qubits are made, a problem
which is beyond the scope of this abstract.

\medskip This work was supported by the Gerard Swope Fund and the Fund
for Encouragement of Research.

%end \small
\end{document}